#  Improved information security using Steganography and Image Segmentation during transmission

Mamta juneja,Parvinder Singh Sandhu

Computer Science and Engineering Department, Rayat and Bahra Institute of Engineering and Technology (RBIEBT)

Sahauran (Punjab), India

er_mamta@yahoo.com

*Abstract*- Steganography is an emerging area which is used for secured data transmission over any public media.Steganography is a process that involves hiding a message in an appropriate carrier like image or audio. It is of Greek origin and means *"covered or hidden writing"*. The carrier can be sent to a receiver without any one except the authenticated receiver knowing the existence of this information. In this paper, a specific image based steganography technique for communicating information more securely between two locations is proposed. The author incorporated the idea of secret key and password security features for authentication at both ends in order to achieve high level of security. As a further improvement of security level, the information has been permuted, encoded and then finally embedded on an image to form the stego image. In addition segmented objects extraction and reassembly of the stego image through normalized cut method has been carried out at the sender side and receiver side respectively in order to prevent distortion of the Stego image during transmission.

Keywords: Steganography, Cover Image, Stego Image, Normalized Cut.

## 1. INTRODUCTION

The term steganography is not new today. In fact several examples from the times of ancient Greece are available in Kahn [3]. In recent years, everything is trending toward digitalization and with the rapid development of the Internet technologies, digital media can be transmitted conveniently over the network. Therefore, messages need to be transmitted secretly through the digital media by using the steganography techniques.Steganography differs from Cryptography in the sense that where cryptography focuses on keeping the contents of a message secret, steganography focuses on keeping the existence of a message secret [6, 15]. Another form of information hiding is digital watermarking, which is the process that embeds data called a watermark, tag or label into a multimedia object such that watermark can be detected or extracted later to make an assertion about the object. The object may be an image, audio, video or text only [9]. Although steganography is an ancient subject, the modern formulation of it comes from the prisoner's problem proposed by Simmons [1], where two prisoners named Alice and Bob wish to communicate in secret to hatch an escape plan. All of their communication passes through a warden named Eve who will throw them solitary confinement if she suspects any type of secret communication. So they must find out someway of hiding their secret message which gives the birth of steganography. The warden is free to examine all communication exchanged between Alice and Bob can either be active or passive. An active warden will try to alter the communication with the suspected hidden information deliberately in order to remove the information where as a passive warden takes the note of covered communication, informs the others and allows the message to pass through .An assumption can be made based on this model is that if both the sender and receiver share some common secret information then the corresponding steganography protocol is known as then the secret key steganography where as pure steganography means that there is none prior information shared by sender and receiver. If the public key of the receiver is known to the sender, the steganographic protocol is called public key steganography [2, 5]. For a more thorough knowledge of steganography methodology the reader may see [7, 15]. Although all digital file formats can be used for steganography, but the image and audio files are more suitable because of their high degree of redundancy [15]. Figure 1 below shows the different categories of file formats that can be used for steganography techniques.

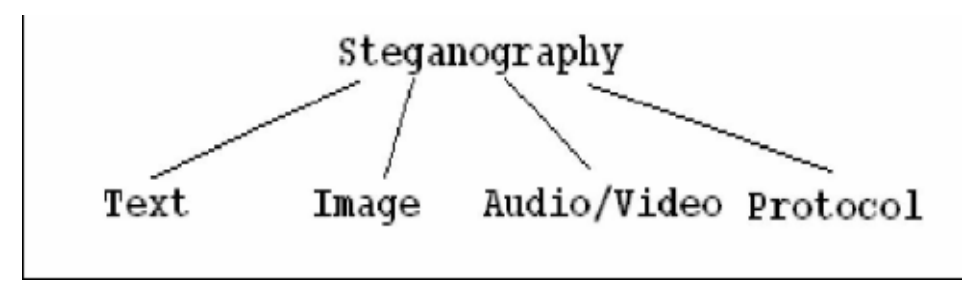

Figure 1: Categories of Steganography

A block diagram of a generic image steganographic system is given in Figure 2. A message is embedded in a digital image (cover image) through an embedding algorithm, with the help of a secret key. The resulting stego image is transmitted over a channel to the receiver where it is processed by the extraction algorithm using the same key. During transmission the stego image, it can be monitored by unauthenticated viewers who will only notice the transmission of an image without discovering the existence of the hidden message. Kevin Curran et al [13] propose an image based steganography methods where he describes a set of steganography methods along with their respective merits and demerits. The most common and simplest image embedding method is the least significant bit (LSB) insertion. The LSB insertion embeds the message in the least significant bit of some selected pixels of the cover image. R.Chadramouli et al. [12] gives an analysis of lsb based steganography techniques. The embedding capacity of LSB method can be increased by using two or more least significant bits. At the same time, not only the risk of making the embedded message statistically detectable increases but also the image fidelity degrades. Hence a variable-sized LSB embedding scheme is presented in [11], in which the number of LSBs used for message embedding /extracting depends on the local characteristics of the pixel. The advantages of LSB-based method are easy to implement. Unfortunately, the hidden message is assailable due to a slight modification from the active warden. Marvel et al. [8] present an image steganographic method, entitled spread spectrum image steganography (SSIS) that hides and recovers the message within digital imagery. The SSIS incorporated the use of error-control codes to correct the large number of bit errors. In recent years many image steganography models have been proposed where the main objective is to protect the transmitted data against any odd. Although increasing the security level of the hidden message of the transmitted data is still an open issue. Silvia Torres Maya et al. [14] presents a steganography algorithm based on bit plane complexity segmentation, which permits to implement hiding information into images for its sure transmission through a non secure channel. In this paper a specific secret-key image based steganographic model proposed which uses an image as the cover data and the secret information is embedded in the cover data to form the stego data which is also an image. The stego image has been divided into several segments using normalized cut method [4,10] and each segments containing the parts of the embedded message transmit separately to the receiver. This work proposes a novel algorithm with higher security features so that the embedded message can not be hacked by unauthorized user.

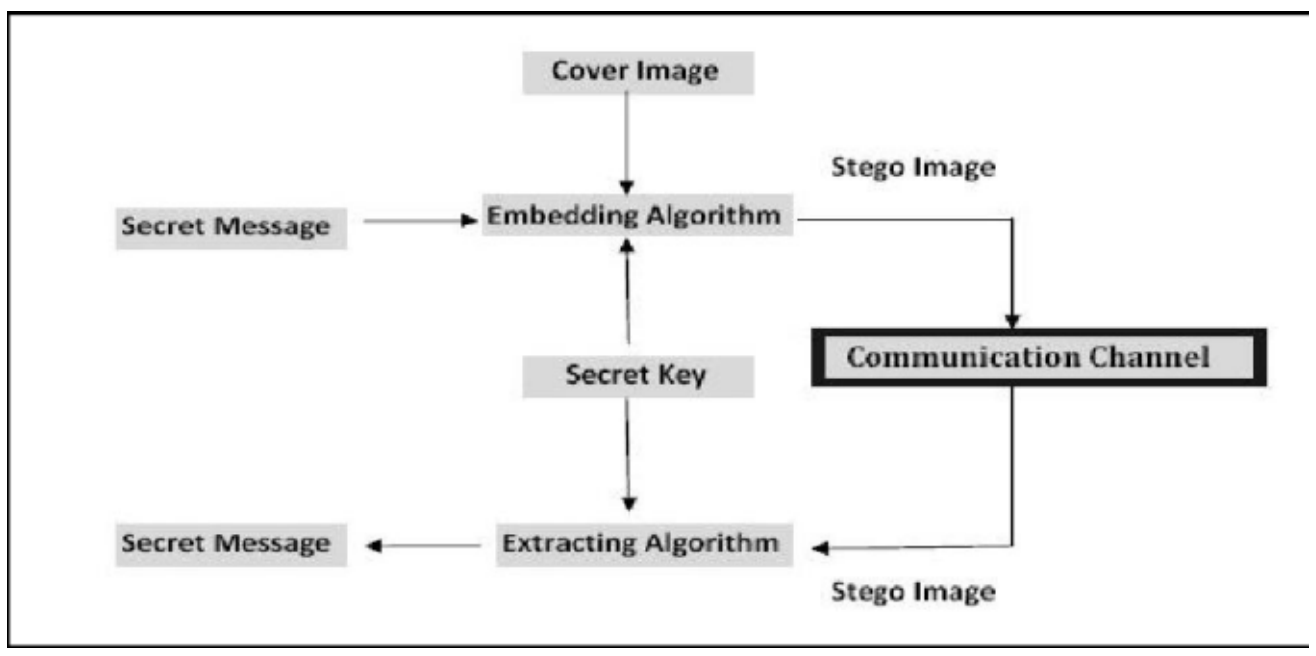

Figure 2: Generic image steganographic system

This paper has been organized as following sections:- Section II deals with proposed model and the solution methodology, Section III contains the analysis of the results and Section IV draws the conclusion.

## 2. THE PROPOSED MODEL

The input messages can be in any digital form, and are often treated as a bit stream. The input message is then converted into encrypted form after a bit permutation is done on the message. Two level of encryption has been done on the message to raise the steganographic security level. This encrypted message generates the secret key, which may be called a message enabled key. A pseudo random number generator is used to locate the embedding positions of the message bits randomly. Before embedding a password protection is also necessary to raise another level of security. Stego image has been segmented using normalized cut method at the sender side and each cut /segmented objects are transmitted individually to the receiver, which are required for reassembly and feature matching at the receiver side. At the receiver side, cuts/segmented objects of the stego image are reassembled and tested for a specific feature. If that feature matches, the extraction process starts after checking the password. The embedded message is extracted but in encrypted form. The extracted message then goes through the consecutive decryption process and re permutation and finally the receiver may be able to see the embedded message with the help of same secret key generated at the sender side .

## 3. ANALYSIS OF THE RESULTS

In the previous work made by different researchers it has been seen that emphasis is given on lsb encoding and decoding technique to make it robust over any image processing operation. In this work an attempt has been made to increase the level of security of the steganography model by incorporating the idea of secret key, password along with the use of permuted and encoded form of the original message. Further the object extraction of the Stego image, reassembly and feature matching has also been used to increase the level of security. The Levels of security incorporated in the proposed model:-

1. Permutation of the secret message.
2. Two level encryption method used in the permuted message.
3. Embedding encrypted form of the message in image.
4. Use of password and secret key.
5. Segmentation, Reassembly of objects and feature matching of the Stego Image.

All the processes both in sender side and receiver side must be executed in proper sequence.

## 4. CONCLUDING REMARKS

The work dealt with the techniques for steganography as related to image science. The results of the test in order to check the security level is quite satisfactory. The security level may be increased combining the approach of text steganography and image steganography. As far as extending this research goes, segmented object extraction method may be substituted with multi-level thresholding and feature based merging to generate optimum objects. The wavelet based image compression over the stego image would be the best candidate to receive further attention.

## 5. REFERENCES

(1) Gustavus J. Simmons, "The Prisoners' Problem and the Subliminal Channel", in Proceedings of CRYPTO '83, pp 51-67. Plenum Press (1984).
(2) "Stretching the Limits of Steganography", RJ Anderson in Information Hiding, Springer Lecture Notes in Computer Science v 1174 (1996) pp 39-48
(3) D. Kahn, The Codebreakers - the comprehensive history of secret communication from ancient times to the Internet, Scribner, New York (1996).
(4) J. Shi and J. Malik, "Normalized Cuts and Image Segmentation," Int. Conf. Computer Vision and Pattern Recognition, San Juan, Puerto Rico, June 1997.
(5) Scott Craver, "On Public-key Steganography in the Presence of an Active Warden," in Proceedings of 2nd International Workshop on Information Hiding, April 1998, Portland, Oregon, USA. pp. 355 - 368.
(6) Ross J. Anderson and Fabien A.P. Petitcolas, "On the limits of steganography," IEEE Journal on Selected Areas in Communications (J-SAC), Special Issue on Copyright & Privacy Protection, vol. 16 no. 4, pp 474-481, May 1998.
(7) N. F. Johnson and S. Jajodia, "Steganography: seeing the unseen," IEEE Computer.,Feb., 26-34 (1998).
(8) L. M. Marvel, C. G. Boncelet, Jr. and C. T. Retter, "Spread spectrum image steganography," IEEE Trans. on Image Processing, 8(8), 1075-1083 (1999).
(9) Digital Watermarking :A Tutorial Review S.P.Mohanty ,1999.
(10) J. Shi and J. Malik, "Normalized cuts and image segmentation.,"IEEE Trans. PAMI, vol. 22, no. 8, pp. 888-905, 2000.
(11) Y. K. Lee and L. H. Chen, "High capacity image steganographic model," IEE Proc.-Vision, Image and Signal Processing, 147(3), 288-294 (2000).
(12) Analysis of LSB Based Image Steganography Techniques ,R. Chandramouli, Nasir Memon, Proc. IEEE ICIP, 2001.

(13) An Evaluation of Image Based Steganography Methods,Kevin Curran, Kran Bailey, International Journal of Digital Evidence,Fall 2003.

(14) Silvia Torres Maya,Mariko Nakano and Ruben Vazquez Medina “Robust Steganography using Bit Plane Complexity Segmentation” 1st International Conference on Electrical and Electronics Engineering ,2004.

(15) T Mrkel,JHP Eloff and MS Olivier .”An Overview of Image Steganography,”in proceedings of the fifth annual Information Security South Africa Conference ,2005.